\documentclass[reprint,aps,prb,showpacs,floatfix,superscriptaddress,citeautoscript,a4paper,longbibliography]{revtex4-1}

\usepackage{amsmath, amsfonts, amssymb, amsbsy}
\usepackage{graphicx}
\usepackage{xspace}

\usepackage{hyperref}
\hypersetup{colorlinks=true, linkcolor=blue, citecolor=blue, urlcolor=blue, unicode=true}

\newcommand*\circled[1]{{\large\textcircled{\small\textsc{#1}}}}

\newcommand{\PbSnSe}{Pb\texorpdfstring{$_{0.73}$}{0.73}Sn\texorpdfstring{$_{0.27}$}{0.27}Se\xspace}
\newcommand{\PbSnXSe}{Pb\texorpdfstring{$_{1-x}$}{\ifpdfstringunicode{\unichar{"2081}\unichar{"208B}\unichar{"2093}}{1-x}}Sn\texorpdfstring{$_{x}$}{\ifpdfstringunicode{\unichar{"2093}}{x}}Se\xspace}
\newcommand{\PbSnXTe}{Pb\texorpdfstring{$_{1-x}$}{\ifpdfstringunicode{\unichar{"2081}\unichar{"208B}\unichar{"2093}}{1-x}}Sn\texorpdfstring{$_{x}$}{\ifpdfstringunicode{\unichar{"2093}}{x}}Te\xspace}

\usepackage{color}

\begin{document}
\title{Spin-polarized (001) surface states of the topological crystalline insulator \PbSnSe}

\author{B.~M. Wojek}
\homepage{http://bastian.wojek.de/}
\affiliation{KTH Royal Institute of Technology, ICT Materials Physics, Electrum 229, 164 40 Kista, Sweden}
\author{R.~Buczko}
\email{buczko@ifpan.edu.pl}
\affiliation{Institute of Physics, Polish Academy of Sciences, Aleja Lotnik\'{o}w 32/46, 02-668 Warsaw, Poland}
\author{S.~Safaei}
\affiliation{Institute of Physics, Polish Academy of Sciences, Aleja Lotnik\'{o}w 32/46, 02-668 Warsaw, Poland}
\author{P.~Dziawa}
\affiliation{Institute of Physics, Polish Academy of Sciences, Aleja Lotnik\'{o}w 32/46, 02-668 Warsaw, Poland}
\author{B.~J. Kowalski}
\affiliation{Institute of Physics, Polish Academy of Sciences, Aleja Lotnik\'{o}w 32/46, 02-668 Warsaw, Poland}
\author{M.~H. Berntsen}
\affiliation{KTH Royal Institute of Technology, ICT Materials Physics, Electrum 229, 164 40 Kista, Sweden}
\author{T.~Balasubramanian}
\affiliation{MAX IV Laboratory, Lund University, P.O. Box 118, 221 00 Lund, Sweden}
\author{M.~Leandersson}
\affiliation{MAX IV Laboratory, Lund University, P.O. Box 118, 221 00 Lund, Sweden}
\author{A.~Szczerbakow}
\affiliation{Institute of Physics, Polish Academy of Sciences, Aleja Lotnik\'{o}w 32/46, 02-668 Warsaw, Poland}
\author{P.~Kacman}
\affiliation{Institute of Physics, Polish Academy of Sciences, Aleja Lotnik\'{o}w 32/46, 02-668 Warsaw, Poland}
\author{T.~Story}
\affiliation{Institute of Physics, Polish Academy of Sciences, Aleja Lotnik\'{o}w 32/46, 02-668 Warsaw, Poland}
\author{O.~Tjernberg}
\affiliation{KTH Royal Institute of Technology, ICT Materials Physics, Electrum 229, 164 40 Kista, Sweden}

\date{\today}

\begin{abstract}
We study the nature of $(001)$ surface states in \PbSnSe in the newly discovered topological-crystalline-insulator (TCI) phase as well as the corresponding topologically trivial state above the band-gap-inversion temperature. Our calculations predict not only metallic surface states with a nontrivial chiral spin structure for the TCI case, but also nonmetallic (gapped) surface states with nonzero spin polarization when the system is a normal insulator. For both phases, angle- and spin-resolved photoelectron spectroscopy measurements provide conclusive evidence for the formation of these $(001)$ surface states in \PbSnSe, as well as for their chiral spin structure.
\end{abstract}


\maketitle

\section{Introduction}
In the recent years, the theoretical prediction~\cite{Kane-PhysRevLett-2005, Fu-PhysRevB-2007, Fu-PhysRevLett-2007} and the successive discovery~\cite{Koenig-Science-2007, Hsieh-Nature-2008, Hsieh-Nature-2009} of topological order in solids have attracted considerable attention shaping the study of topological phenomena into one of the major fields of contemporary condensed-matter physics~\cite{Hasan-RevModPhys-2010, Qi-RevModPhys-2011}. In the case of topological insulators (TIs), bulk semiconductors with an odd number of band inversions support gapless Dirac-like surface states. These are topologically protected by time-reversal symmetry and feature a chiral spin texture, thus providing robust spin-polarized conduction channels. Naturally, it is this property that renders TIs potential candidate materials for spintronics applications~\cite{Yazyev-PhysRevLett-2010, Pesin-NatMater-2012}.

In the search for new TIs, lately, a novel topologically nontrivial state has been proposed to exist in SnTe~\cite{Hsieh-NatCommun-2012}. Despite not falling in the class of the ``classic'' TIs with $Z_2$ topological invariants~\cite{Fu-PhysRevB-2007}, this IV-VI narrow-gap semiconductor hosts metallic surface states protected by the mirror symmetry of the crystal's rock-salt (RS) structure---hence the name: \emph{topological crystalline insulator} (TCI)~\cite{Hsieh-NatCommun-2012}. More specifically, this state is identified with the $fm\overline{3}m-f3(4)$ valley phase in an extended classification scheme newly put forward~\cite{Slager-arXiv-2012}. It has been shown by angle-resolved photoelectron spectroscopy (ARPES) studies that this state is indeed realized on the $(001)$ surfaces of SnTe~\cite{Tanaka-NatPhys-2012}, as well as of the IV-VI substitutional alloys \PbSnXTe~\cite{Xu-NatCommun-2012} and \PbSnXSe~\cite{Dziawa-NatMater-2012}.

Although, in principle the spin-orbit interaction (SOI) is not a prerequisite for obtaining a TCI~\cite{Fu-PhysRevLett-2011}, the TCI phase in real SnTe- or SnSe-based compounds would not be achieved without the relativistic effects (SOI and Darwin term). Due to the SOI, the metallic surface states observed in these materials have almost linear, Dirac-like, dispersions~\cite{Hsieh-NatCommun-2012, Tanaka-NatPhys-2012, Xu-NatCommun-2012, Dziawa-NatMater-2012}. Moreover, spin-polarization effects are anticipated for these states~\cite{Hsieh-NatCommun-2012, Xu-NatCommun-2012}.

In this article, we present a theoretical prediction for the spin polarization of metallic $(001)$ surface states in the TCI phase of \PbSnSe, i.e., when the band structure is inverted at low temperatures. Surprisingly enough, at high temperatures, when the system is a normal insulator, a nonzero spin polarization of gapped surface states is obtained in the calculations. Experimental evidence for the formation of the surface states and their chiral spin texture in \PbSnSe single crystals at both, low and high temperatures is provided through ARPES and spin-resolved photoelectron spectroscopy (SRPES). 

\section{Band-structure calculations}
To determine the electronic surface states in \PbSnSe we have performed tight-binding (TB) calculations for a slab with $162$ atomic monolayers. Such TB modeling proved to be very efficient for describing mixed crystals and their heterostructures, e.g., \PbSnXTe, Pb$_{1-x}$Cd$_{x}$Te, and Pb$_{1-x}$Mn$_{x}$Te alloys~\cite{Dziawa-NatMater-2012, Bukala-PhysRevB-2012, Bukala-NanoscaleResLett-2011}. The slab has a RS crystal structure and is oriented in $[001]$ direction. In the two directions parallel to the surface periodic boundary conditions have been imposed. To describe the RS substitutional alloy within the virtual crystal approximation the TB parameters for the constituent compounds, PbSe and SnSe, are needed. While TB parameters for PbSe can be taken from the literature~\cite{Lent-SuperlattMicrostruct-1986}, the parameters for RS SnSe are not available, as SnSe crystallizes in an orthorhombic structure. To obtain the parameters for the hypothetical material, bulk RS SnSe, we first estimate its 
electronic band structure using standard \emph{ab-initio} methods, i.e., density-functional theory with local-density approximation (DFT-LDA). The only experimental result, which can be used to verify the obtained DFT band structure of RS SnSe is the linear 
change of the band gap of \PbSnXSe with the Sn content---at $T=4$~K the band gap $E_{\mathrm{g}}$ changes from a value of $0.165$~eV for $x=0$, via $E_{\mathrm{g}} = 0$ for $x\approx0.19$ to $E_{\mathrm{g}} = -0.129$~eV for $x\approx0.33$~\cite{Strauss-PhysRev-1967}. Thus, the final TB parameters for SnSe are obtained from the parametrization of the DFT-LDA band structure by use of the simulated annealing method~\cite{Ingber-MathComputModell-1989} with the additional condition of a linear change of $E_{\mathrm{g}}(x)$ in the mixed crystal. In each step of this procedure, the parameters of the \PbSnXSe alloy are obtained by linear interpolation of the actual TB parameters of PbSe and SnSe. In our TB Hamiltonian for SnSe the $s$, $p$ and $d$ orbitals and nearest-neighbor interactions are included. Finally, by extrapolating the measured $E_{\mathrm{g}}(x)$ from Ref.~\onlinecite{Strauss-PhysRev-1967} to $x=1$, one finds the ``experimental'' band gap of RS SnSe to be $-0.725$~eV while the energy difference 
between the DFT-LDA conduction and valence bands in the $L$ point is equal to $-0.981$~eV. We rescale the obtained parameters for SnSe to obtain the ``experimental'' band gap and, therefore, the proper $E_{\mathrm{g}}(x)$ dependence for the \PbSnXSe alloy, in particular with the energy-gap sign change occurring at $x=0.19$.

In our experiment the change of $E_{\mathrm{g}}$ is achieved not by changing the Sn content but by varying the temperature. Thus, for a direct comparison of the experimental findings with the theory we have to check how the TB parameters of \PbSnSe evolve with temperature. We assume that this dependence is dominated by the change of the lattice constant $a_0$ with temperature and rescale again our parametrization according to the Harrison rules~\cite{Harrison}. We assume that $a_0$ for any Sn content diminishes by the same amount with decreasing temperature as for PbSe~\cite{Preier-ApplPhys-1979}. The room-temperature $a_0$ for \PbSnSe is taken from X-ray-diffraction experiments~\cite{Szczerbakow-JCrystalGrowth-1994}. With the procedure described above we recover exactly the experimental temperature dependence of the band gap for PbSe. With increasing Sn content the agreement diminishes, but in the interesting region of $x=0.27$, the resulting $E_{\mathrm{g}}(T)$ differs from that given in Ref.~\onlinecite{
Strauss-PhysRev-1967} by less than about $15$~\%.

\begin{figure*}
\centering
\includegraphics[width=.8\textwidth]{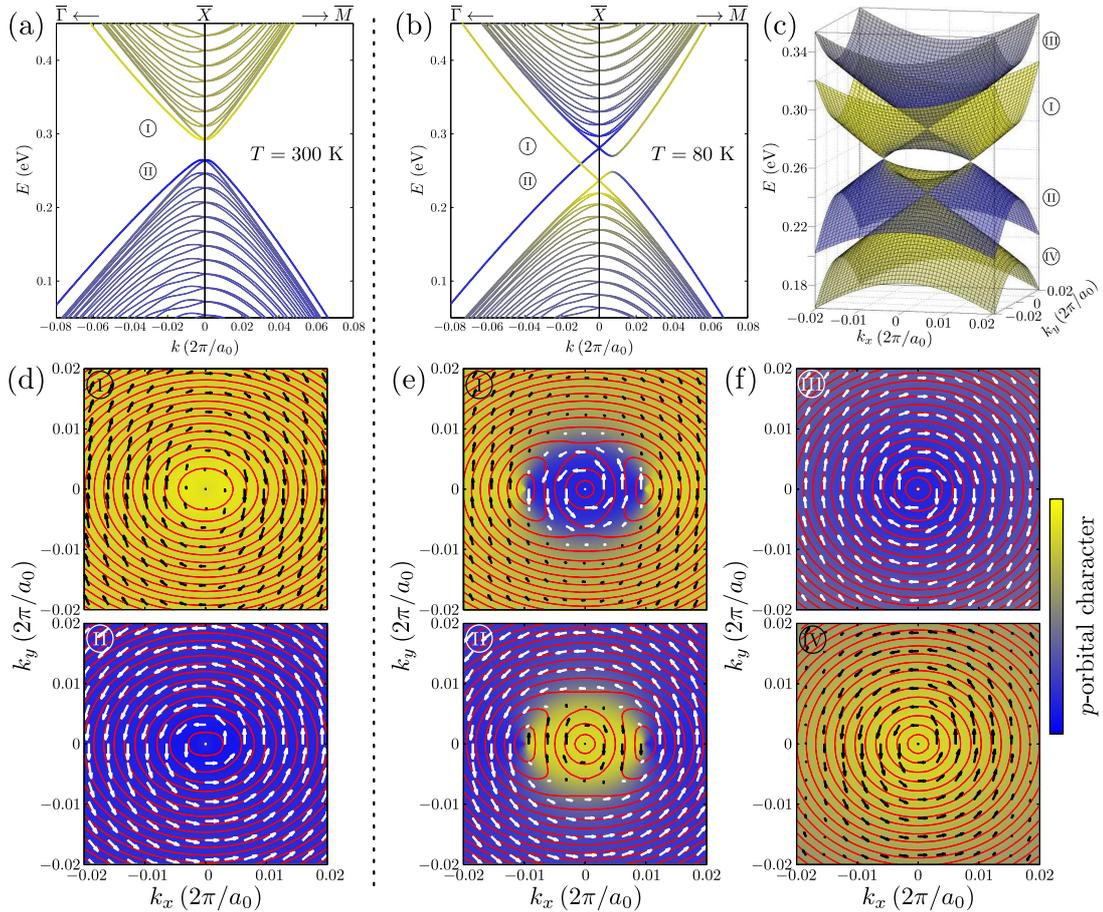}
\caption{(a) Band structure of the \PbSnSe $(001)$ slab in the vicinity of $\overline{X}$ at $T=300$~K ($a_0=6.091$~\AA) along the high-symmetry lines. The $\overline{X}$ position was taken as $(k_x,k_y)=(0,0)$. The surface states near the conduction and valence bands are denoted by \circled{i} and \circled{ii}, respectively. (b) The same band structure at $T=80$~K ($a_0 = 6.066$~\AA). Here only one Dirac point (where the surface states \circled{i} and \circled{ii} are in contact) is shown on the $\overline{\Gamma}$-$\overline{X}$ line. Both Dirac points are visible in (c), where only the four bands with energies within the bulk-band-gap region are presented. (d) and (e) are contour plots of the constant-energy lines of the surface states labeled by \circled{i} and \circled{ii} at the two studied temperatures. The arrows indicate the in-plane spin texture, their size the degree of spin polarization. (f) shows the same for the next bands \circled{iii} and \circled{iv}. The blue to yellow color coding 
indicates the contributions of the cation (yellow) and anion (blue) $p$ orbitals to the wavefunctions.}
\label{fig:PbSnSe-theory}
\end{figure*}

As shown in Fig.~\ref{fig:PbSnSe-theory}(a), the band calculations for the \PbSnSe $(001)$ surface at $T=300$~K yield a noninverted band structure and hence no gapless states. Yet, the lowest (\circled{i}) and highest (\circled{ii}) states of the conduction and valence bands, respectively, represent states localized at the surface. Figures~\ref{fig:PbSnSe-theory}(b) and~(c) show the band structure of \PbSnSe close to $\overline{X}$ in the TCI state. Analogous to the situation in SnTe~\cite{Hsieh-NatCommun-2012}, surface states are found to cross the inverted bulk band gap along the $\overline{\Gamma}$-$\overline{X}$ direction. The surface states above (\circled{i} and \circled{iii}) and below (\circled{ii} and \circled{iv}) the Dirac points are depicted in Fig.~\ref{fig:PbSnSe-theory}(c). The calculated spin textures for the states \circled{i} and \circled{ii} are presented in Fig.~\ref{fig:PbSnSe-theory} (d) for $300$~K and~(e) for $80$~K. As can be seen in the figure, these states are spin-polarized and 
have a chiral spin texture in both cases. We note that the clockwise (counter-clockwise) chirality is related to the cation (anion) $p$-type orbitals. Also, while at high temperature single vortices are anticipated, the spin patterns presented in Fig.~\ref{fig:PbSnSe-theory}(e) for low temperature consist of double-vortical structures. In the latter, for the band \circled{i} and wave vectors $\mathbf{k}$ between the Dirac points (inner vortex) the spin rotates counter-clockwise about $\overline{X}$. Outside this region (outer vortex) the rotation is reversed (clockwise). Thus, in the local environment of each Dirac point the same left-handed chiral structure is obtained, like predicted in Ref.~\onlinecite{Hsieh-NatCommun-2012}. For energies below the Dirac point, i.e., for band \circled{ii}, all the chiralities are reversed as compared to band \circled{i}. Moreover, it is seen that the polarization is expected to be slightly larger for the anions than for the cations. In contrast, the spin textures of the 
higher (\circled{iii}) and lower (\circled{iv}) bands in Fig.~\ref{fig:PbSnSe-theory}(f), have the form of single vortices. This results from the two additional Dirac points at $\overline{X}$. In one of these, the bands \circled{i} and \circled{iii} are brought to contact forming anion Dirac cones, while the bands \circled{ii} and \circled{iv} meet in the other to form cation Dirac cones [cf. Fig.~\ref{fig:PbSnSe-theory}(b) and~(c)]. These Dirac points are not topologically protected, but still have an impact on the overall spin structure. The spins in the anion (cation) Dirac cones of the \circled{iii} (\circled{iv}) bands are rotating in the opposite direction as compared to the anion (cation) spins in the \circled{i} (\circled{ii}) bands.
 
\section{Photoemission experiments}

\begin{figure*}
\includegraphics[width=.9\textwidth]{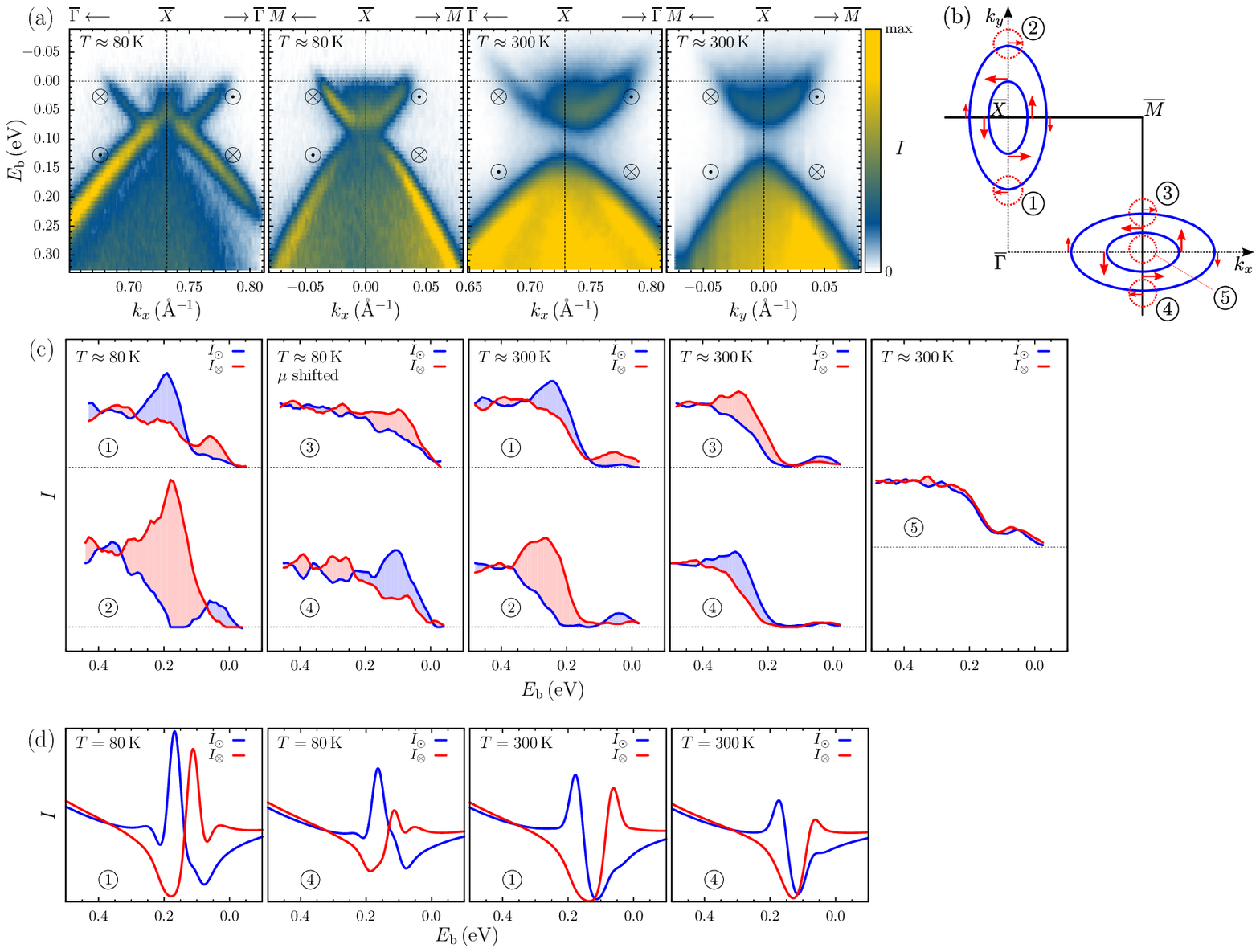}
\caption{(a) ARPES spectra of \PbSnSe along the indicated high-symmetry lines through $\overline{X}$ in the TCI phase ($T\approx 80$~K) and for the topologically trivial state ($T\approx 300$~K). The data have been obtained from adjacent sides of the $(001)$ surface Brillouin zone [see sketch in (b)]. \begin{Large}$\odot$\end{Large} and \begin{Large}$\otimes$\end{Large} show the spin polarization for the respective surface-state branches. (b) Sketch of the first quadrant of the $(001)$ surface Brillouin zone of \PbSnSe. The blue ellipses indicate surface-state constant-energy lines between the conduction-band minimum along $\overline{X}$-$\overline{M}$ and the anion Dirac point at $\overline{X}$ in the TCI state (\emph{not to scale}). The red arrows show the corresponding electron spin texture [cf. Fig.~\ref{fig:PbSnSe-theory}(e)]. The red dotted circles labeled by the numbers \circled{1} through \circled{5} give the rough positions for the SRPES measurements of the in-plane spin component approximately 
parallel to $k_{x}$. (c) SRPES spectra for the five reciprocal-space points indicated in (b). The absolute intensities shown on a linear ordinate scale where the dotted lines indicate zero, were obtained using Eq.~(\ref{eq:absoluteIntensity}). Note that the chemical potential $\mu$ is shifted in the measurements at point \circled{3} and \circled{4} at $T\approx 80$~K, so that the states above the Dirac points (at lower binding energies) hardly contribute to the spectra. (d) Theoretically expected absolute intensities (without $\mu$ cutoff).}
\label{fig:PES}
\end{figure*}

To test the results of our calculations we conducted photoemission experiments on single crystals of \PbSnSe. The studied $n$-type single crystals have been grown by the self-selecting vapor growth method under near-equilibrium thermodynamic conditions~\cite{Szczerbakow-JCrystalGrowth-1994, Szczerbakow-ProgCrystalGrowth-2005}. The ARPES and SRPES measurements on cleaved $(001)$ surfaces at temperatures $T\approx80$~K and $T\approx300$~K have been carried out at the I3 and I4 beam lines at the MAX-III synchrotron at MAX-lab, Lund University, Sweden~\cite{Berntsen-RevSciInstrum-2010, Jensen-NIMA-1997}. All experiments have been done under ultra-high-vacuum conditions ($p\approx 3\times10^{-10}$~mbar) using linearly polarized light with a photon energy $h\nu=18.5$~eV. The total energy resolution of the ARPES and SRPES experiments was about ($10$~to~$20$)~meV and $100$~meV, respectively. The in-plane crystal-momentum resolution was $0.01$~\AA{}$^{-1}$ for the ARPES measurements. The SRPES data acquisition 
integrated over a reciprocal-space area with a diameter of about $0.025$~\AA{}$^{-1}$.

Figure~\ref{fig:PES}(a) shows ARPES spectra for both sample temperatures along the high-symmetry lines of the $(001)$ Brillouin zone ($\overline{\Gamma}$-$\overline{X}$-$\overline{\Gamma}$ and $\overline{M}$-$\overline{X}$-$\overline{M}$) in the vicinity of $\overline{X}$. In agreement with previous ARPES measurements on Pb$_{0.77}$Sn$_{0.23}$Se~\cite{Dziawa-NatMater-2012}, the formation of gapless surface states along $\overline{\Gamma}$-$\overline{X}$-$\overline{\Gamma}$ characteristic for the TCI phase takes place at low temperatures, whereas at high temperatures, only gapped states are observed. The band-structure inversion in \PbSnSe occurs at $T\approx250$~K~\cite{Strauss-PhysRev-1967}, thus, the material is either a TCI ($T\approx80$~K) or a normal insulator ($T\approx300$~K). While our calculations for the room-temperature case indicate a paraboloid-like dispersion centered at $\overline{X}$, the corresponding data suggest that the conduction-band minima are shifted away from $\overline{X}$ [cf. $\overline{\Gamma}$-$\overline{X}$-$\overline{\Gamma}$ dispersion in Fig.~\ref{fig:PES}(a)].

On the basis of these ARPES results, successively, the in-plane spin texture was probed by SRPES measurements employing the mini-Mott polarimeter at the I3 beam line~\cite{Berntsen-RevSciInstrum-2010}. This set-up is used to determine the projection of the in-plane electron spin on a given direction---in this case approximately the direction parallel to $k_x$ as sketched in Fig.~\ref{fig:PES}(c). In the process, two photoemission spectra [energy-distribution curves (EDCs)] for electrons with opposite spins are acquired and the spin polarization $\mathcal{P}(E)$ is given by the asymmetry
\begin{equation}
 S_{\mathrm{eff}}\,\mathcal{P}(E)=\frac{\hat{I}_{\otimes}(E)-\hat{I}_{\odot}(E)}{\hat{I}_{\otimes}(E)+\hat{I}_{\odot}(E)}.
\label{eq:polarization}
\end{equation}
Here, $\hat{I}_{\otimes,\odot}(E)$ are the background-corrected and normalized intensities in the two spectra and $S_{\mathrm{eff}}\approx0.2$ is the effective Sherman function, a measure for the efficiency of the Mott detector~\cite{Gay-RevSciInstrum-1992}. Eventually, the absolute photoemission intensities $I_{\otimes,\odot}(E)$ corrected for $S_{\mathrm{eff}} < 1$ are obtained by
\begin{equation}
 I_{\genfrac{}{}{0pt}{2}{\otimes}{\odot}}(E)=\frac{\hat{I}_{\otimes}(E)+\hat{I}_{\odot}(E)}{2}\left[1\pm\mathcal{P}(E)\right].
\label{eq:absoluteIntensity}
\end{equation}

Unfortunately, during the low-temperature measurements, adsorption of residual gases in the vacuum chamber and reactions with the photon beam cause a rapid shift of the chemical potential of the samples of up to $100$~meV/$30$~min, thus successively emptying the surface states. Hence, a quantitative polarization analysis based on high-statistics measurements is unfeasible. Nevertheless, initial SRPES measurements could be performed at the five reciprocal-space points indicated in Fig.~\ref{fig:PES}(b). Figure~\ref{fig:PES}(c) summarizes their results. The measured spectra have been normalized to yield on average no net spin polarization for the contributions to the spectra from bulk-like states at binding energies higher than $350$~meV. To improve the clarity of the presentation a simple moving average of three measured data points (corresponding to an energy interval of $20$~meV) has been taken to calculate the shown spectra. Figure~\ref{fig:PES}(d) shows the expected intensities at the different $k$-space 
points based on our calculations. It has been assumed that the probability of photon absorption is the same for all states and that the photoemission probability decays exponentially (with a decay constant of $0.7$~nm) with the distance from the surface. An additional Gaussian energy broadening of $20$~meV has been applied.

Altogether, the clear intensity asymmetry for photoelectrons with opposite spins indicates a finite spin polarization at both temperatures. Although the experimentally and theoretically obtained dispersion curves at $T=300$~K somewhat seem to deviate, the observed spin polarizations agree qualitatively. The states have a chiral spin texture which reverses when crossing the Dirac-point energy or the band gap, respectively. The polarization is given mainly by the surface states for both temperatures. The calculated small but nonzero polarization of the bulk states can be understood by recalling that only the spin polarization of the first few monolayers closest to the surface is considered. Since the surface breaks the inversion symmetry, the states in this region may be polarized. Of course, the total spin polarization for the doubly degenerate bulk-like states of the slab with two surfaces is equal to zero.

A comparison of the data with the calculations in Fig.~\ref{fig:PbSnSe-theory}(e) shows that in the TCI state the anticipated ``outer'' spin rotation in the bands \circled{i} and \circled{ii} is realized. On the other hand, the identification of the ``inner'' spin rotation remains elusive. Any attempt to measure the polarization at a position between one of the Dirac points and $\overline{X}$ is hindered by the inevitable averaging over a considerably large $k$-space region, the overall lower photoemission intensity from the surface states [cf. Fig.~\ref{fig:PES}(a)] and at low temperatures the additional shift of the chemical potential. Since the Dirac points are much further apart in the related system Pb$_{0.6}$Sn$_{0.4}$Te, the observation of the ``inner'' spin structure in the TCI state of this material occurred to be possible~\cite{Xu-NatCommun-2012}.

\section{Discussion and Conclusions}
The experimentally observed chiral spin texture of the metallic surface states in the TCI phase of \PbSnSe is consistent with our theoretical TB model as well as with the very recent report on the spin texture in the TCI phase of the closely related Pb$_{0.6}$Sn$_{0.4}$Te alloy~\cite{Xu-NatCommun-2012}. Here we have shown that in the normal-insulator state of \PbSnSe gapped spin-polarized surface states exist as well.
IV-VI semiconductors and their substitutional alloys crystallize in the RS structure and possess global inversion symmetry. Therefore, only structure-inversion-asymmetry-induced spin splitting (Rashba effect)~\cite{Winkler-SOI, Zawadzki-SemicondSciTechnol-2004} is expected for the two-dimensional surface states in these semiconductors. It has been checked that within our TB model the calculated dispersion at high temperatures tends to a single paraboloid centered at $\overline{X}$ when the crystal slab thickness increases. On the other hand, our ARPES data suggest the presence of split bands. The latter might be related to phenomena not included in the calculations, e.g., the Rashba effect given by the confining potential at the charged surface. Yet, clearly both, more elaborate experimental as well as theoretical work is needed to provide further details for these spin-polarized gapped surface states.

To summarize, the spin polarization of the $(001)$ surface states have been predicted in both, the newly discovered TCI phase as well as in the topologically trivial state of \PbSnSe. While our calculations show a ``double-vortical'' spin structure for the TCI state, in the normal-insulator state a ``single-vortical'' spin texture is anticipated. ARPES measurements confirm the formation of the TCI phase below, but also indicate split gapped states above the band-gap-inversion temperature. SRPES provides conclusive evidence for the realization of a chiral spin texture in both cases, overall consistent with the calculations. As revealed by our studies, the spin polarization seems to be inherent to surface states in narrow-gap IV-VI semiconductors influenced by very strong SOI. The transition to the TCI phase brought about by the band-symmetry inversion qualitatively influences orbital degrees of freedom of electrons and holes (cation/anion wavefunction inversion). This, via SOI, results in the spin polarization of electrons occupying both, in-gap Dirac-metal surface states as well as the states close to the bottom of the conduction band and the top of the valence band whose detailed nature, however, still awaits further elucidation.

\acknowledgments
This work was made possible through support from the Knut and Alice Wallenberg Foundation, the Swedish Research Council, the European Commission Network SemiSpinNet (PITN-GA-2008-215368), the European Regional Development Fund through the Innovative Economy grant (POIG.01.01.02-00-108/09), and the Polish National Science Centre (NCN) Grant No. 2011/03/B/ST3/02659. P.~D. and B.~J.~K. acknowledge the support from the Baltic Science Link project coordinated by the Swedish Research Council, VR.


\begin{thebibliography}{30}%
\makeatletter
\providecommand \@ifxundefined [1]{%
 \@ifx{#1\undefined}
}%
\providecommand \@ifnum [1]{%
 \ifnum #1\expandafter \@firstoftwo
 \else \expandafter \@secondoftwo
 \fi
}%
\providecommand \@ifx [1]{%
 \ifx #1\expandafter \@firstoftwo
 \else \expandafter \@secondoftwo
 \fi
}%
\providecommand \natexlab [1]{#1}%
\providecommand \enquote  [1]{``#1''}%
\providecommand \bibnamefont  [1]{#1}%
\providecommand \bibfnamefont [1]{#1}%
\providecommand \citenamefont [1]{#1}%
\providecommand \href@noop [0]{\@secondoftwo}%
\providecommand \href [0]{\begingroup \@sanitize@url \@href}%
\providecommand \@href[1]{\@@startlink{#1}\@@href}%
\providecommand \@@href[1]{\endgroup#1\@@endlink}%
\providecommand \@sanitize@url [0]{\catcode `\\12\catcode `\$12\catcode
  `\&12\catcode `\#12\catcode `\^12\catcode `\_12\catcode `\%12\relax}%
\providecommand \@@startlink[1]{}%
\providecommand \@@endlink[0]{}%
\providecommand \url  [0]{\begingroup\@sanitize@url \@url }%
\providecommand \@url [1]{\endgroup\@href {#1}{\urlprefix }}%
\providecommand \urlprefix  [0]{URL }%
\providecommand \Eprint [0]{\href }%
\providecommand \doibase [0]{http://dx.doi.org/}%
\providecommand \selectlanguage [0]{\@gobble}%
\providecommand \bibinfo  [0]{\@secondoftwo}%
\providecommand \bibfield  [0]{\@secondoftwo}%
\providecommand \translation [1]{[#1]}%
\providecommand \BibitemOpen [0]{}%
\providecommand \bibitemStop [0]{}%
\providecommand \bibitemNoStop [0]{.\EOS\space}%
\providecommand \EOS [0]{\spacefactor3000\relax}%
\providecommand \BibitemShut  [1]{\csname bibitem#1\endcsname}%
\let\auto@bib@innerbib\@empty
\bibitem [{\citenamefont {Kane}\ and\ \citenamefont
  {Mele}(2005)}]{Kane-PhysRevLett-2005}%
  \BibitemOpen
  \bibfield  {author} {\bibinfo {author} {\bibfnamefont {C.~L.}\ \bibnamefont
  {Kane}}\ and\ \bibinfo {author} {\bibfnamefont {E.~J.}\ \bibnamefont
  {Mele}},\ }\bibfield  {title} {\enquote {\bibinfo {title} {{${Z}_{2}$
  Topological Order and the Quantum Spin Hall Effect}},}\ }\href {\doibase
  10.1103/PhysRevLett.95.146802} {\bibfield  {journal} {\bibinfo  {journal}
  {Phys. Rev. Lett.}\ }\textbf {\bibinfo {volume} {95}},\ \bibinfo {pages}
  {146802} (\bibinfo {year} {2005})}\BibitemShut {NoStop}%
\bibitem [{\citenamefont {Fu}\ and\ \citenamefont
  {Kane}(2007)}]{Fu-PhysRevB-2007}%
  \BibitemOpen
  \bibfield  {author} {\bibinfo {author} {\bibfnamefont {Liang}\ \bibnamefont
  {Fu}}\ and\ \bibinfo {author} {\bibfnamefont {C.~L.}\ \bibnamefont {Kane}},\
  }\bibfield  {title} {\enquote {\bibinfo {title} {Topological insulators with
  inversion symmetry},}\ }\href {\doibase 10.1103/PhysRevB.76.045302}
  {\bibfield  {journal} {\bibinfo  {journal} {Phys. Rev. B}\ }\textbf {\bibinfo
  {volume} {76}},\ \bibinfo {pages} {045302} (\bibinfo {year}
  {2007})}\BibitemShut {NoStop}%
\bibitem [{\citenamefont {Fu}\ \emph {et~al.}(2007)\citenamefont {Fu},
  \citenamefont {Kane},\ and\ \citenamefont {Mele}}]{Fu-PhysRevLett-2007}%
  \BibitemOpen
  \bibfield  {author} {\bibinfo {author} {\bibfnamefont {Liang}\ \bibnamefont
  {Fu}}, \bibinfo {author} {\bibfnamefont {C.~L.}\ \bibnamefont {Kane}}, \ and\
  \bibinfo {author} {\bibfnamefont {E.~J.}\ \bibnamefont {Mele}},\ }\bibfield
  {title} {\enquote {\bibinfo {title} {Topological insulators in three
  dimensions},}\ }\href {\doibase 10.1103/PhysRevLett.98.106803} {\bibfield
  {journal} {\bibinfo  {journal} {Phys. Rev. Lett.}\ }\textbf {\bibinfo
  {volume} {98}},\ \bibinfo {pages} {106803} (\bibinfo {year}
  {2007})}\BibitemShut {NoStop}%
\bibitem [{\citenamefont {K\"onig}\ \emph {et~al.}(2007)\citenamefont
  {K\"onig}, \citenamefont {Wiedmann}, \citenamefont {Br\"une}, \citenamefont
  {Roth}, \citenamefont {Buhmann}, \citenamefont {Molenkamp}, \citenamefont
  {Qi},\ and\ \citenamefont {Zhang}}]{Koenig-Science-2007}%
  \BibitemOpen
  \bibfield  {author} {\bibinfo {author} {\bibfnamefont {Markus}\ \bibnamefont
  {K\"onig}}, \bibinfo {author} {\bibfnamefont {Steffen}\ \bibnamefont
  {Wiedmann}}, \bibinfo {author} {\bibfnamefont {Christoph}\ \bibnamefont
  {Br\"une}}, \bibinfo {author} {\bibfnamefont {Andreas}\ \bibnamefont {Roth}},
  \bibinfo {author} {\bibfnamefont {Hartmut}\ \bibnamefont {Buhmann}}, \bibinfo
  {author} {\bibfnamefont {Laurens~W.}\ \bibnamefont {Molenkamp}}, \bibinfo
  {author} {\bibfnamefont {Xiao-Liang}\ \bibnamefont {Qi}}, \ and\ \bibinfo
  {author} {\bibfnamefont {Shou-Cheng}\ \bibnamefont {Zhang}},\ }\bibfield
  {title} {\enquote {\bibinfo {title} {{Quantum Spin Hall Insulator State in
  HgTe Quantum Wells}},}\ }\href {\doibase 10.1126/science.1148047} {\bibfield
  {journal} {\bibinfo  {journal} {Science}\ }\textbf {\bibinfo {volume}
  {318}},\ \bibinfo {pages} {766--770} (\bibinfo {year} {2007})}\BibitemShut
  {NoStop}%
\bibitem [{\citenamefont {Hsieh}\ \emph {et~al.}(2008)\citenamefont {Hsieh},
  \citenamefont {Qian}, \citenamefont {Wray}, \citenamefont {Xia},
  \citenamefont {Hor}, \citenamefont {Cava},\ and\ \citenamefont
  {Hasan}}]{Hsieh-Nature-2008}%
  \BibitemOpen
  \bibfield  {author} {\bibinfo {author} {\bibfnamefont {D.}~\bibnamefont
  {Hsieh}}, \bibinfo {author} {\bibfnamefont {D.}~\bibnamefont {Qian}},
  \bibinfo {author} {\bibfnamefont {L.}~\bibnamefont {Wray}}, \bibinfo {author}
  {\bibfnamefont {Y.}~\bibnamefont {Xia}}, \bibinfo {author} {\bibfnamefont
  {Y.~S.}\ \bibnamefont {Hor}}, \bibinfo {author} {\bibfnamefont {R.~J.}\
  \bibnamefont {Cava}}, \ and\ \bibinfo {author} {\bibfnamefont {M.~Z.}\
  \bibnamefont {Hasan}},\ }\bibfield  {title} {\enquote {\bibinfo {title} {{A
  topological Dirac insulator in a quantum spin Hall phase}},}\ }\href
  {\doibase 10.1038/nature06843} {\bibfield  {journal} {\bibinfo  {journal}
  {Nature}\ }\textbf {\bibinfo {volume} {452}},\ \bibinfo {pages} {970--974}
  (\bibinfo {year} {2008})}\BibitemShut {NoStop}%
\bibitem [{\citenamefont {Hsieh}\ \emph {et~al.}(2009)\citenamefont {Hsieh},
  \citenamefont {Xia}, \citenamefont {Qian}, \citenamefont {Wray},
  \citenamefont {Dil}, \citenamefont {Meier}, \citenamefont {Osterwalder},
  \citenamefont {Patthey}, \citenamefont {Checkelsky}, \citenamefont {Ong},
  \citenamefont {Fedorov}, \citenamefont {Lin}, \citenamefont {Bansil},
  \citenamefont {Grauer}, \citenamefont {Hor}, \citenamefont {Cava},\ and\
  \citenamefont {Hasan}}]{Hsieh-Nature-2009}%
  \BibitemOpen
  \bibfield  {author} {\bibinfo {author} {\bibfnamefont {D.}~\bibnamefont
  {Hsieh}}, \bibinfo {author} {\bibfnamefont {Y.}~\bibnamefont {Xia}}, \bibinfo
  {author} {\bibfnamefont {D.}~\bibnamefont {Qian}}, \bibinfo {author}
  {\bibfnamefont {L.}~\bibnamefont {Wray}}, \bibinfo {author} {\bibfnamefont
  {J.~H.}\ \bibnamefont {Dil}}, \bibinfo {author} {\bibfnamefont
  {F.}~\bibnamefont {Meier}}, \bibinfo {author} {\bibfnamefont
  {J.}~\bibnamefont {Osterwalder}}, \bibinfo {author} {\bibfnamefont
  {L.}~\bibnamefont {Patthey}}, \bibinfo {author} {\bibfnamefont {J.~G.}\
  \bibnamefont {Checkelsky}}, \bibinfo {author} {\bibfnamefont {N.~P.}\
  \bibnamefont {Ong}}, \bibinfo {author} {\bibfnamefont {A.~V.}\ \bibnamefont
  {Fedorov}}, \bibinfo {author} {\bibfnamefont {H.}~\bibnamefont {Lin}},
  \bibinfo {author} {\bibfnamefont {A.}~\bibnamefont {Bansil}}, \bibinfo
  {author} {\bibfnamefont {D.}~\bibnamefont {Grauer}}, \bibinfo {author}
  {\bibfnamefont {Y.~S.}\ \bibnamefont {Hor}}, \bibinfo {author} {\bibfnamefont
  {R.~J.}\ \bibnamefont {Cava}}, \ and\ \bibinfo {author} {\bibfnamefont
  {M.~Z.}\ \bibnamefont {Hasan}},\ }\bibfield  {title} {\enquote {\bibinfo
  {title} {{A tunable topological insulator in the spin helical Dirac transport
  regime}},}\ }\href {\doibase 10.1038/nature08234} {\bibfield  {journal}
  {\bibinfo  {journal} {Nature}\ }\textbf {\bibinfo {volume} {460}},\ \bibinfo
  {pages} {1101--1105} (\bibinfo {year} {2009})}\BibitemShut {NoStop}%
\bibitem [{\citenamefont {Hasan}\ and\ \citenamefont
  {Kane}(2010)}]{Hasan-RevModPhys-2010}%
  \BibitemOpen
  \bibfield  {author} {\bibinfo {author} {\bibfnamefont {M.~Z.}\ \bibnamefont
  {Hasan}}\ and\ \bibinfo {author} {\bibfnamefont {C.~L.}\ \bibnamefont
  {Kane}},\ }\bibfield  {title} {\enquote {\bibinfo {title}
  {{\textit{Colloquium}: Topological insulators}},}\ }\href {\doibase
  10.1103/RevModPhys.82.3045} {\bibfield  {journal} {\bibinfo  {journal} {Rev.
  Mod. Phys.}\ }\textbf {\bibinfo {volume} {82}},\ \bibinfo {pages}
  {3045--3067} (\bibinfo {year} {2010})}\BibitemShut {NoStop}%
\bibitem [{\citenamefont {Qi}\ and\ \citenamefont
  {Zhang}(2011)}]{Qi-RevModPhys-2011}%
  \BibitemOpen
  \bibfield  {author} {\bibinfo {author} {\bibfnamefont {Xiao-Liang}\
  \bibnamefont {Qi}}\ and\ \bibinfo {author} {\bibfnamefont {Shou-Cheng}\
  \bibnamefont {Zhang}},\ }\bibfield  {title} {\enquote {\bibinfo {title}
  {Topological insulators and superconductors},}\ }\href {\doibase
  10.1103/RevModPhys.83.1057} {\bibfield  {journal} {\bibinfo  {journal} {Rev.
  Mod. Phys.}\ }\textbf {\bibinfo {volume} {83}},\ \bibinfo {pages}
  {1057--1110} (\bibinfo {year} {2011})}\BibitemShut {NoStop}%
\bibitem [{\citenamefont {Yazyev}\ \emph {et~al.}(2010)\citenamefont {Yazyev},
  \citenamefont {Moore},\ and\ \citenamefont
  {Louie}}]{Yazyev-PhysRevLett-2010}%
  \BibitemOpen
  \bibfield  {author} {\bibinfo {author} {\bibfnamefont {Oleg~V.}\ \bibnamefont
  {Yazyev}}, \bibinfo {author} {\bibfnamefont {Joel~E.}\ \bibnamefont {Moore}},
  \ and\ \bibinfo {author} {\bibfnamefont {Steven~G.}\ \bibnamefont {Louie}},\
  }\bibfield  {title} {\enquote {\bibinfo {title} {{Spin Polarization and
  Transport of Surface States in the Topological Insulators
  ${\mathrm{Bi}}_{2}{\mathrm{Se}}_{3}$ and ${\mathrm{Bi}}_{2}{\mathrm{Te}}_{3}$
  from First Principles}},}\ }\href {\doibase 10.1103/PhysRevLett.105.266806}
  {\bibfield  {journal} {\bibinfo  {journal} {Phys. Rev. Lett.}\ }\textbf
  {\bibinfo {volume} {105}},\ \bibinfo {pages} {266806} (\bibinfo {year}
  {2010})}\BibitemShut {NoStop}%
\bibitem [{\citenamefont {Pesin}\ and\ \citenamefont
  {MacDonald}(2012)}]{Pesin-NatMater-2012}%
  \BibitemOpen
  \bibfield  {author} {\bibinfo {author} {\bibfnamefont {Dmytro}\ \bibnamefont
  {Pesin}}\ and\ \bibinfo {author} {\bibfnamefont {Allan~H.}\ \bibnamefont
  {MacDonald}},\ }\bibfield  {title} {\enquote {\bibinfo {title} {{Spintronics
  and pseudospintronics in graphene and topological insulators}},}\ }\href
  {\doibase 10.1038/nmat3305} {\bibfield  {journal} {\bibinfo  {journal} {Nat.
  Mater.}\ }\textbf {\bibinfo {volume} {11}},\ \bibinfo {pages} {409--416}
  (\bibinfo {year} {2012})}\BibitemShut {NoStop}%
\bibitem [{\citenamefont {Hsieh}\ \emph {et~al.}(2012)\citenamefont {Hsieh},
  \citenamefont {Lin}, \citenamefont {Liu}, \citenamefont {Duan}, \citenamefont
  {Bansil},\ and\ \citenamefont {Fu}}]{Hsieh-NatCommun-2012}%
  \BibitemOpen
  \bibfield  {author} {\bibinfo {author} {\bibfnamefont {Timothy~H.}\
  \bibnamefont {Hsieh}}, \bibinfo {author} {\bibfnamefont {Hsin}\ \bibnamefont
  {Lin}}, \bibinfo {author} {\bibfnamefont {Junwei}\ \bibnamefont {Liu}},
  \bibinfo {author} {\bibfnamefont {Wenhui}\ \bibnamefont {Duan}}, \bibinfo
  {author} {\bibfnamefont {Arun}\ \bibnamefont {Bansil}}, \ and\ \bibinfo
  {author} {\bibfnamefont {Liang}\ \bibnamefont {Fu}},\ }\bibfield  {title}
  {\enquote {\bibinfo {title} {{Topological crystalline insulators in the SnTe
  material class}},}\ }\href {\doibase 10.1038/ncomms1969} {\bibfield
  {journal} {\bibinfo  {journal} {Nat. Commun.}\ }\textbf {\bibinfo {volume}
  {3}},\ \bibinfo {pages} {982} (\bibinfo {year} {2012})}\BibitemShut {NoStop}%
\bibitem [{\citenamefont {Slager}\ \emph {et~al.}(2012)\citenamefont {Slager},
  \citenamefont {Mesaros}, \citenamefont {Juri\v{c}i\'{c}},\ and\ \citenamefont
  {Zaanen}}]{Slager-arXiv-2012}%
  \BibitemOpen
  \bibfield  {author} {\bibinfo {author} {\bibfnamefont {Robert-Jan}\
  \bibnamefont {Slager}}, \bibinfo {author} {\bibfnamefont {Andrej}\
  \bibnamefont {Mesaros}}, \bibinfo {author} {\bibfnamefont {Vladimir}\
  \bibnamefont {Juri\v{c}i\'{c}}}, \ and\ \bibinfo {author} {\bibfnamefont
  {Jan}\ \bibnamefont {Zaanen}},\ }\href@noop {} {\enquote {\bibinfo {title}
  {{The space group classification of topological band insulators}},}\ }
  (\bibinfo {year} {2012}),\ \Eprint {http://arxiv.org/abs/1209.2610}
  {arXiv:1209.2610} \BibitemShut {NoStop}%
\bibitem [{\citenamefont {Tanaka}\ \emph {et~al.}(2012)\citenamefont {Tanaka},
  \citenamefont {Ren}, \citenamefont {Sato}, \citenamefont {Nakayama},
  \citenamefont {Souma}, \citenamefont {Takahashi}, \citenamefont {Segawa},\
  and\ \citenamefont {Ando}}]{Tanaka-NatPhys-2012}%
  \BibitemOpen
  \bibfield  {author} {\bibinfo {author} {\bibfnamefont {Y.}~\bibnamefont
  {Tanaka}}, \bibinfo {author} {\bibfnamefont {Zhi}\ \bibnamefont {Ren}},
  \bibinfo {author} {\bibfnamefont {T.}~\bibnamefont {Sato}}, \bibinfo {author}
  {\bibfnamefont {K.}~\bibnamefont {Nakayama}}, \bibinfo {author}
  {\bibfnamefont {S.}~\bibnamefont {Souma}}, \bibinfo {author} {\bibfnamefont
  {T.}~\bibnamefont {Takahashi}}, \bibinfo {author} {\bibfnamefont {Kouji}\
  \bibnamefont {Segawa}}, \ and\ \bibinfo {author} {\bibfnamefont {Yoichi}\
  \bibnamefont {Ando}},\ }\bibfield  {title} {\enquote {\bibinfo {title}
  {{Experimental realization of a topological crystalline insulator in
  SnTe}},}\ }\href {\doibase 10.1038/nphys2442} {\bibfield  {journal} {\bibinfo
   {journal} {Nat. Phys.}\ }\textbf {\bibinfo {volume} {8}},\ \bibinfo {pages}
  {800--803} (\bibinfo {year} {2012})}\BibitemShut {NoStop}%
\bibitem [{\citenamefont {Xu}\ \emph {et~al.}(2012)\citenamefont {Xu},
  \citenamefont {Liu}, \citenamefont {Alidoust}, \citenamefont {Neupane},
  \citenamefont {Qian}, \citenamefont {Belopolski}, \citenamefont {Denlinger},
  \citenamefont {Wang}, \citenamefont {Lin}, \citenamefont {Wray},
  \citenamefont {Landolt}, \citenamefont {Slomski}, \citenamefont {Dil},
  \citenamefont {Marcinkova}, \citenamefont {Morosan}, \citenamefont {Gibson},
  \citenamefont {Sankar}, \citenamefont {Chou}, \citenamefont {Cava},
  \citenamefont {Bansil},\ and\ \citenamefont {Hasan}}]{Xu-NatCommun-2012}%
  \BibitemOpen
  \bibfield  {author} {\bibinfo {author} {\bibfnamefont {Su-Yang}\ \bibnamefont
  {Xu}}, \bibinfo {author} {\bibfnamefont {Chang}\ \bibnamefont {Liu}},
  \bibinfo {author} {\bibfnamefont {N.}~\bibnamefont {Alidoust}}, \bibinfo
  {author} {\bibfnamefont {M.}~\bibnamefont {Neupane}}, \bibinfo {author}
  {\bibfnamefont {D.}~\bibnamefont {Qian}}, \bibinfo {author} {\bibfnamefont
  {I.}~\bibnamefont {Belopolski}}, \bibinfo {author} {\bibfnamefont {J.~D.}\
  \bibnamefont {Denlinger}}, \bibinfo {author} {\bibfnamefont {Y.~J.}\
  \bibnamefont {Wang}}, \bibinfo {author} {\bibfnamefont {H.}~\bibnamefont
  {Lin}}, \bibinfo {author} {\bibfnamefont {L.~A.}\ \bibnamefont {Wray}},
  \bibinfo {author} {\bibfnamefont {G.}~\bibnamefont {Landolt}}, \bibinfo
  {author} {\bibfnamefont {B.}~\bibnamefont {Slomski}}, \bibinfo {author}
  {\bibfnamefont {J.~H.}\ \bibnamefont {Dil}}, \bibinfo {author} {\bibfnamefont
  {A.}~\bibnamefont {Marcinkova}}, \bibinfo {author} {\bibfnamefont
  {E.}~\bibnamefont {Morosan}}, \bibinfo {author} {\bibfnamefont
  {Q.}~\bibnamefont {Gibson}}, \bibinfo {author} {\bibfnamefont
  {R.}~\bibnamefont {Sankar}}, \bibinfo {author} {\bibfnamefont {F.~C.}\
  \bibnamefont {Chou}}, \bibinfo {author} {\bibfnamefont {R.~J.}\ \bibnamefont
  {Cava}}, \bibinfo {author} {\bibfnamefont {A.}~\bibnamefont {Bansil}}, \ and\
  \bibinfo {author} {\bibfnamefont {M.~Z.}\ \bibnamefont {Hasan}},\ }\bibfield
  {title} {\enquote {\bibinfo {title} {{Observation of a topological
  crystalline insulator phase and topological phase transition in
  Pb$_{1-x}$Sn$_x$Te}},}\ }\href {\doibase 10.1038/ncomms2191} {\bibfield
  {journal} {\bibinfo  {journal} {Nat. Commun.}\ }\textbf {\bibinfo {volume}
  {3}},\ \bibinfo {pages} {1192} (\bibinfo {year} {2012})}\BibitemShut
  {NoStop}%
\bibitem [{\citenamefont {Dziawa}\ \emph {et~al.}(2012)\citenamefont {Dziawa},
  \citenamefont {Kowalski}, \citenamefont {Dybko}, \citenamefont {Szczerbakow},
  \citenamefont {Szot}, \citenamefont {\L{}usakowska}, \citenamefont
  {Balasubramanian}, \citenamefont {Wojek}, \citenamefont {Berntsen},
  \citenamefont {Tjernberg},\ and\ \citenamefont
  {Story}}]{Dziawa-NatMater-2012}%
  \BibitemOpen
  \bibfield  {author} {\bibinfo {author} {\bibfnamefont {P.}~\bibnamefont
  {Dziawa}}, \bibinfo {author} {\bibfnamefont {B.~J.}\ \bibnamefont
  {Kowalski}}, \bibinfo {author} {\bibfnamefont {K.}~\bibnamefont {Dybko}},
  \bibinfo {author} {\bibfnamefont {A.}~\bibnamefont {Szczerbakow}}, \bibinfo
  {author} {\bibfnamefont {M.}~\bibnamefont {Szot}}, \bibinfo {author}
  {\bibfnamefont {E.}~\bibnamefont {\L{}usakowska}}, \bibinfo {author}
  {\bibfnamefont {T.}~\bibnamefont {Balasubramanian}}, \bibinfo {author}
  {\bibfnamefont {B.~M.}\ \bibnamefont {Wojek}}, \bibinfo {author}
  {\bibfnamefont {M.~H.}\ \bibnamefont {Berntsen}}, \bibinfo {author}
  {\bibfnamefont {O.}~\bibnamefont {Tjernberg}}, \ and\ \bibinfo {author}
  {\bibfnamefont {T.}~\bibnamefont {Story}},\ }\bibfield  {title} {\enquote
  {\bibinfo {title} {{Topological crystalline insulator states in
  Pb$_{1-x}$Sn$_{x}$Se}},}\ }\href {\doibase 10.1038/nmat3449} {\bibfield
  {journal} {\bibinfo  {journal} {Nat. Mater.}\ }\textbf {\bibinfo {volume}
  {11}},\ \bibinfo {pages} {1023--1027} (\bibinfo {year} {2012})}\BibitemShut
  {NoStop}%
\bibitem [{\citenamefont {Fu}(2011)}]{Fu-PhysRevLett-2011}%
  \BibitemOpen
  \bibfield  {author} {\bibinfo {author} {\bibfnamefont {Liang}\ \bibnamefont
  {Fu}},\ }\bibfield  {title} {\enquote {\bibinfo {title} {Topological
  crystalline insulators},}\ }\href {\doibase 10.1103/PhysRevLett.106.106802}
  {\bibfield  {journal} {\bibinfo  {journal} {Phys. Rev. Lett.}\ }\textbf
  {\bibinfo {volume} {106}},\ \bibinfo {pages} {106802} (\bibinfo {year}
  {2011})}\BibitemShut {NoStop}%
\bibitem [{\citenamefont {Buka\l{}a}\ \emph {et~al.}(2012)\citenamefont
  {Buka\l{}a}, \citenamefont {Sankowski}, \citenamefont {Buczko},\ and\
  \citenamefont {Kacman}}]{Bukala-PhysRevB-2012}%
  \BibitemOpen
  \bibfield  {author} {\bibinfo {author} {\bibfnamefont {M.}~\bibnamefont
  {Buka\l{}a}}, \bibinfo {author} {\bibfnamefont {P.}~\bibnamefont
  {Sankowski}}, \bibinfo {author} {\bibfnamefont {R.}~\bibnamefont {Buczko}}, \
  and\ \bibinfo {author} {\bibfnamefont {P.}~\bibnamefont {Kacman}},\
  }\bibfield  {title} {\enquote {\bibinfo {title} {{Structural and electronic
  properties of Pb$_{1-x}$Cd$_{x}$Te and Pb$_{1-x}$Mn$_{x}$Te ternary
  alloys}},}\ }\href {\doibase 10.1103/PhysRevB.86.085205} {\bibfield
  {journal} {\bibinfo  {journal} {Phys. Rev. B}\ }\textbf {\bibinfo {volume}
  {86}},\ \bibinfo {pages} {085205} (\bibinfo {year} {2012})}\BibitemShut
  {NoStop}%
\bibitem [{\citenamefont {Buka\l{}a}\ \emph {et~al.}(2011)\citenamefont
  {Buka\l{}a}, \citenamefont {Sankowski}, \citenamefont {Buczko},\ and\
  \citenamefont {Kacman}}]{Bukala-NanoscaleResLett-2011}%
  \BibitemOpen
  \bibfield  {author} {\bibinfo {author} {\bibfnamefont {Ma\l{}gorzata}\
  \bibnamefont {Buka\l{}a}}, \bibinfo {author} {\bibfnamefont {Piotr}\
  \bibnamefont {Sankowski}}, \bibinfo {author} {\bibfnamefont {Ryszard}\
  \bibnamefont {Buczko}}, \ and\ \bibinfo {author} {\bibfnamefont {Per\l{}a}\
  \bibnamefont {Kacman}},\ }\bibfield  {title} {\enquote {\bibinfo {title}
  {{Crystal and electronic structure of PbTe/CdTe nanostructures}},}\ }\href
  {\doibase 10.1186/1556-276X-6-126} {\bibfield  {journal} {\bibinfo  {journal}
  {Nanoscale Res. Lett.}\ }\textbf {\bibinfo {volume} {6}},\ \bibinfo {pages}
  {126} (\bibinfo {year} {2011})}\BibitemShut {NoStop}%
\bibitem [{\citenamefont {Lent}\ \emph {et~al.}(1986)\citenamefont {Lent},
  \citenamefont {Bowen}, \citenamefont {Dow}, \citenamefont {Allgaier},
  \citenamefont {Sankey},\ and\ \citenamefont
  {Ho}}]{Lent-SuperlattMicrostruct-1986}%
  \BibitemOpen
  \bibfield  {author} {\bibinfo {author} {\bibfnamefont {Craig~S.}\
  \bibnamefont {Lent}}, \bibinfo {author} {\bibfnamefont {Marshall~A.}\
  \bibnamefont {Bowen}}, \bibinfo {author} {\bibfnamefont {John~D.}\
  \bibnamefont {Dow}}, \bibinfo {author} {\bibfnamefont {Robert~S.}\
  \bibnamefont {Allgaier}}, \bibinfo {author} {\bibfnamefont {Otto~F.}\
  \bibnamefont {Sankey}}, \ and\ \bibinfo {author} {\bibfnamefont {Eliza~S.}\
  \bibnamefont {Ho}},\ }\bibfield  {title} {\enquote {\bibinfo {title}
  {{Relativistic empirical tight-binding theory of the energy bands of GeTe,
  SnTe, PbTe, PbSe, PbS, and their alloys}},}\ }\href {\doibase
  10.1016/0749-6036(86)90017-0} {\bibfield  {journal} {\bibinfo  {journal}
  {Superlattices Microstruct.}\ }\textbf {\bibinfo {volume} {2}},\ \bibinfo
  {pages} {491--499} (\bibinfo {year} {1986})}\BibitemShut {NoStop}%
\bibitem [{\citenamefont {Strauss}(1967)}]{Strauss-PhysRev-1967}%
  \BibitemOpen
  \bibfield  {author} {\bibinfo {author} {\bibfnamefont {A.~J.}\ \bibnamefont
  {Strauss}},\ }\bibfield  {title} {\enquote {\bibinfo {title} {{Inversion of
  Conduction and Valence Bands in
  ${\mathrm{Pb}}_{1-x}{\mathrm{Sn}}_{x}\mathrm{Se}$ Alloys}},}\ }\href
  {\doibase 10.1103/PhysRev.157.608} {\bibfield  {journal} {\bibinfo  {journal}
  {Phys. Rev.}\ }\textbf {\bibinfo {volume} {157}},\ \bibinfo {pages}
  {608--611} (\bibinfo {year} {1967})}\BibitemShut {NoStop}%
\bibitem [{\citenamefont {Ingber}(1989)}]{Ingber-MathComputModell-1989}%
  \BibitemOpen
  \bibfield  {author} {\bibinfo {author} {\bibfnamefont {L.}~\bibnamefont
  {Ingber}},\ }\bibfield  {title} {\enquote {\bibinfo {title} {Very fast
  simulated re-annealing},}\ }\href {\doibase 10.1016/0895-7177(89)90202-1}
  {\bibfield  {journal} {\bibinfo  {journal} {Math. Comput. Modell.}\ }\textbf
  {\bibinfo {volume} {12}},\ \bibinfo {pages} {967--973} (\bibinfo {year}
  {1989})}\BibitemShut {NoStop}%
\bibitem [{\citenamefont {Harrison}(1980)}]{Harrison}%
  \BibitemOpen
  \bibfield  {author} {\bibinfo {author} {\bibfnamefont {Walter~A.}\
  \bibnamefont {Harrison}},\ }\href@noop {} {\emph {\bibinfo {title}
  {{Electronic Structure and the Properties of Solids---The Physics of the
  Chemical Bond}}}}\ (\bibinfo  {publisher} {Freeman},\ \bibinfo {address} {San
  Francisco},\ \bibinfo {year} {1980})\BibitemShut {NoStop}%
\bibitem [{\citenamefont {Preier}(1979)}]{Preier-ApplPhys-1979}%
  \BibitemOpen
  \bibfield  {author} {\bibinfo {author} {\bibfnamefont {H.}~\bibnamefont
  {Preier}},\ }\bibfield  {title} {\enquote {\bibinfo {title} {Recent advances
  in lead-chalcogenide diode lasers},}\ }\href {\doibase 10.1007/BF00886018}
  {\bibfield  {journal} {\bibinfo  {journal} {Appl. Phys.}\ }\textbf {\bibinfo
  {volume} {20}},\ \bibinfo {pages} {189--206} (\bibinfo {year}
  {1979})}\BibitemShut {NoStop}%
\bibitem [{\citenamefont {Szczerbakow}\ and\ \citenamefont
  {Berger}(1994)}]{Szczerbakow-JCrystalGrowth-1994}%
  \BibitemOpen
  \bibfield  {author} {\bibinfo {author} {\bibfnamefont {Andrzej}\ \bibnamefont
  {Szczerbakow}}\ and\ \bibinfo {author} {\bibfnamefont {Hans}\ \bibnamefont
  {Berger}},\ }\bibfield  {title} {\enquote {\bibinfo {title} {{Investigation
  of the composition of vapour-grown Pb$_{1-x}$Sn$_{x}$Se crystals ($x \le
  0.4$) by means of lattice parameter measurements}},}\ }\href {\doibase
  10.1016/0022-0248(94)90042-6} {\bibfield  {journal} {\bibinfo  {journal} {J.
  Cryst. Growth}\ }\textbf {\bibinfo {volume} {139}},\ \bibinfo {pages}
  {172--178} (\bibinfo {year} {1994})}\BibitemShut {NoStop}%
\bibitem [{\citenamefont {Szczerbakow}\ and\ \citenamefont
  {Durose}(2005)}]{Szczerbakow-ProgCrystalGrowth-2005}%
  \BibitemOpen
  \bibfield  {author} {\bibinfo {author} {\bibfnamefont {Andrzej}\ \bibnamefont
  {Szczerbakow}}\ and\ \bibinfo {author} {\bibfnamefont {Ken}\ \bibnamefont
  {Durose}},\ }\bibfield  {title} {\enquote {\bibinfo {title} {{Self-selecting
  vapour growth of bulk crystals -- Principles and applicability}},}\ }\href
  {\doibase 10.1016/j.pcrysgrow.2005.10.004} {\bibfield  {journal} {\bibinfo
  {journal} {Prog. Cryst. Growth Charact. Mater.}\ }\textbf {\bibinfo {volume}
  {51}},\ \bibinfo {pages} {81--108} (\bibinfo {year} {2005})}\BibitemShut
  {NoStop}%
\bibitem [{\citenamefont {Berntsen}\ \emph {et~al.}(2010)\citenamefont
  {Berntsen}, \citenamefont {Palmgren}, \citenamefont {Leandersson},
  \citenamefont {Hahlin}, \citenamefont {\AA{}hlund}, \citenamefont {Wannberg},
  \citenamefont {M\aa{}nsson},\ and\ \citenamefont
  {Tjernberg}}]{Berntsen-RevSciInstrum-2010}%
  \BibitemOpen
  \bibfield  {author} {\bibinfo {author} {\bibfnamefont {M.~H.}\ \bibnamefont
  {Berntsen}}, \bibinfo {author} {\bibfnamefont {P.}~\bibnamefont {Palmgren}},
  \bibinfo {author} {\bibfnamefont {M.}~\bibnamefont {Leandersson}}, \bibinfo
  {author} {\bibfnamefont {A.}~\bibnamefont {Hahlin}}, \bibinfo {author}
  {\bibfnamefont {J.}~\bibnamefont {\AA{}hlund}}, \bibinfo {author}
  {\bibfnamefont {B.}~\bibnamefont {Wannberg}}, \bibinfo {author}
  {\bibfnamefont {M.}~\bibnamefont {M\aa{}nsson}}, \ and\ \bibinfo {author}
  {\bibfnamefont {O.}~\bibnamefont {Tjernberg}},\ }\bibfield  {title} {\enquote
  {\bibinfo {title} {A spin- and angle-resolving photoelectron spectrometer},}\
  }\href {\doibase 10.1063/1.3342120} {\bibfield  {journal} {\bibinfo
  {journal} {Rev. Sci. Instrum.}\ }\textbf {\bibinfo {volume} {81}},\ \bibinfo
  {eid} {035104} (\bibinfo {year} {2010})}\BibitemShut {NoStop}%
\bibitem [{\citenamefont {Jensen}\ \emph {et~al.}(1997)\citenamefont {Jensen},
  \citenamefont {Butorin}, \citenamefont {Kaurila}, \citenamefont {Nyholm},\
  and\ \citenamefont {Johansson}}]{Jensen-NIMA-1997}%
  \BibitemOpen
  \bibfield  {author} {\bibinfo {author} {\bibfnamefont {B.~N.}\ \bibnamefont
  {Jensen}}, \bibinfo {author} {\bibfnamefont {S.~M.}\ \bibnamefont {Butorin}},
  \bibinfo {author} {\bibfnamefont {T.}~\bibnamefont {Kaurila}}, \bibinfo
  {author} {\bibfnamefont {R.}~\bibnamefont {Nyholm}}, \ and\ \bibinfo {author}
  {\bibfnamefont {L.~I.}\ \bibnamefont {Johansson}},\ }\bibfield  {title}
  {\enquote {\bibinfo {title} {Design and performance of a spherical grating
  monochromator used at {MAX I}},}\ }\href {\doibase
  10.1016/S0168-9002(97)00595-0} {\bibfield  {journal} {\bibinfo  {journal}
  {Nucl. Instrum. Methods Phys. Res., Sect. A}\ }\textbf {\bibinfo {volume}
  {394}},\ \bibinfo {pages} {243--250} (\bibinfo {year} {1997})}\BibitemShut
  {NoStop}%
\bibitem [{\citenamefont {Gay}\ and\ \citenamefont
  {Dunning}(1992)}]{Gay-RevSciInstrum-1992}%
  \BibitemOpen
  \bibfield  {author} {\bibinfo {author} {\bibfnamefont {T.~J.}\ \bibnamefont
  {Gay}}\ and\ \bibinfo {author} {\bibfnamefont {F.~B.}\ \bibnamefont
  {Dunning}},\ }\bibfield  {title} {\enquote {\bibinfo {title} {Mott electron
  polarimetry},}\ }\href {\doibase 10.1063/1.1143371} {\bibfield  {journal}
  {\bibinfo  {journal} {Rev. Sci. Instrum.}\ }\textbf {\bibinfo {volume}
  {63}},\ \bibinfo {pages} {1635--1651} (\bibinfo {year} {1992})}\BibitemShut
  {NoStop}%
\bibitem [{\citenamefont {Winkler}(2003)}]{Winkler-SOI}%
  \BibitemOpen
  \bibfield  {author} {\bibinfo {author} {\bibfnamefont {Roland}\ \bibnamefont
  {Winkler}},\ }\href {\doibase 10.1007/b13586} {\emph {\bibinfo {title}
  {{Spin-Orbit Coupling Effects in Two-Dimensional Electron and Hole
  Systems}}}},\ \bibinfo {series} {Springer Tracts in Modern Physics}, Vol.\
  \bibinfo {volume} {191}\ (\bibinfo  {publisher} {Springer},\ \bibinfo
  {address} {Berlin Heidelberg},\ \bibinfo {year} {2003})\BibitemShut {NoStop}%
\bibitem [{\citenamefont {Zawadzki}\ and\ \citenamefont
  {Pfeffer}(2004)}]{Zawadzki-SemicondSciTechnol-2004}%
  \BibitemOpen
  \bibfield  {author} {\bibinfo {author} {\bibfnamefont {W.}~\bibnamefont
  {Zawadzki}}\ and\ \bibinfo {author} {\bibfnamefont {P.}~\bibnamefont
  {Pfeffer}},\ }\bibfield  {title} {\enquote {\bibinfo {title} {Spin splitting
  of subband energies due to inversion asymmetry in semiconductor
  heterostructures},}\ }\href {\doibase 10.1088/0268-1242/19/1/R01} {\bibfield
  {journal} {\bibinfo  {journal} {Semicond. Sci. Technol.}\ }\textbf {\bibinfo
  {volume} {19}},\ \bibinfo {pages} {R1} (\bibinfo {year} {2004})}\BibitemShut
  {NoStop}%
\end{thebibliography}
%

\end{document}